# Increasing situational awareness through nowcasting of the reproduction number

Andrea Bizzotto [a,b,*], Giorgio Guzzetta [a,*], Valentina Marziano [a], Martina del Manso [c], Alberto Mateo Urdiales [c], Daniele Petrone [c], Andrea Cannone [c], Chiara Sacco [c], Piero Poletti [a], Mattia Manica [a], Agnese Zardini [a], Filippo Trentini [d,e], Massimo Fabiani [c], Antonino Bella [c], Flavia Riccardo [c], Patrizio Pezzotti [c,#], Marco Ajelli [f,#], Stefano Merler [a,#,&]

[a] Center for Health Emergencies, Bruno Kessler Foundation, Trento, Italy
[b] University of Trento, Italy
[c] Department of Infectious Diseases, Istituto Superiore di Sanità, Rome, Italy
[d] Covid Crisis Lab, Bocconi University, Milan, Italy
[e] Department of Social and Political Sciences, Bocconi University, Milan, Italy
[f] Laboratory for Computational Epidemiology and Public Health, Department of Epidemiology and Biostatistics, Indiana University School of Public Health, Bloomington, Indiana, United States

[*] Joint first authors
[#] Joint senior authors
[&] Corresponding author:

Stefano Merler – merler@fbk.eu
Center for Health Emergencies
Fondazione Bruno Kessler
Via Sommarive 18, 38123
Povo (Trento), Italy
+390461314595

## Abstract

The time varying reproduction number R is a critical variable for situational awareness during infectious disease outbreaks, but delays between infection and reporting hinder its accurate estimation in real time.

We propose a nowcasting method for improving the timeliness and accuracy of R estimates, based on comparisons of successive versions of surveillance databases. The method was validated against COVID-19 surveillance data collected in Italy over an 18-month period.

Compared to traditional methods, the nowcasted reproduction number reduced the estimation delay from 13 to 8 days, while maintaining a better accuracy. Moreover, it allowed anticipating the detection of periods of epidemic growth by between 6 and 23 days.

The method offers a simple and generally applicable tool to improve situational awareness during an epidemic outbreak, allowing for informed public health response planning.

**First author biosketch**: Andrea Bizzotto is a PhD student in Mathematics at the University of Trento. His research is focused on the development of mathematical and statistical models to investigate the transmission of emerging and re-emerging pathogens in human populations, with public health applications.

## Introduction

Epidemiological surveillance is a critical tool for policy making, allowing public health professionals to monitor epidemic trends and the effectiveness of the adopted interventions. One important quantity that can be monitored during an epidemic outbreak by relying on surveillance system data is the time-varying reproduction number ($R$). The reproduction number is defined as the average number of secondary infections caused by an average infectious individual and represents a summary metric that measures changes in transmissibility over time, indicating whether and how fast an epidemic is growing (when $R > 1$) or declining (when $R < 1$) [1,2]. $R$ can be estimated with established statistical methods [1,3,4,5] from the time series of the number of cases occurring in a given geographic unit (also known as "epidemic curve"), provided that the generation time distribution of the considered infection is known.

Surveillance systems are in most cases unable to trace the date at which cases were infected, and the temporally closest proxy event that can be measured is the onset of symptoms. Therefore, estimating $R$ from epidemic curves organized as number of cases by date of symptom onset provides the closest estimate in time to the actual transmission events, even when symptomatic cases represent a small subset of all identified cases [6]. Indeed, it has been shown that estimates of the reproduction number are robust as long as the proportion between symptomatic cases and total infections remains stable or even drifts slowly over time [5,7].

Between the onset of symptoms for a patient and the insertion of their record in the surveillance database, temporal delays occur depending on the medical-seeking behavior of the individual, the logistics of case ascertainment (e.g., testing), the organization of healthcare response systems (e.g., administration of epidemiological questionnaires, contact tracing), and the socio-technological infrastructure for data collection, quality control, data upload and integration. These delays may change over time not only as a function of the progressive improvement of organizational aspects as the outbreak develops but also depending on the saturation of resources for diagnosis, contact tracing, data collection and transmission, and on other factors such as the population's perceived importance of medical seeking at different stages of the epidemic. As much as surveillance systems can be optimized to minimize some components of these delays, their data provide information that is always somewhat lagged with respect to the current epidemiological situation. As a result of the above-mentioned delays, the epidemic curves will usually be underestimated for symptom onset dates close to the date of reporting and will consolidate only in successive data updates. This eventually limits the ability of public health officers to promptly assess the current situation or the effectiveness of recently implemented interventions. Methods for adjusting incidence curves with respect to reporting delays have been previously proposed in the wake of the AIDS pandemic [8–12], and one of them has been applied as a proof of concept on COVID-19 data [13]. The underestimation of incidence curves in recent dates makes the following two questions of critical importance when estimating $R$ for real-time monitoring purposes: i) until what date in the past can the epidemic curve (and therefore $R$ estimates) be trusted, and ii) how the incompleteness in recent data affects the accuracy of $R$ estimates.

In this study, we propose a method for estimating consolidation delays as well as improving the timeliness and accuracy of the $R$ estimate by adjusting incidence curves. The method is benchmarked against Italian COVID-19 surveillance data collected over more than 18 months, but it is readily generalizable to other infectious diseases.

## Methods

<u>Definitions</u>

Given a generic infectious disease, we define $D$ as a generic date of update of the epidemic curve by the epidemiological surveillance system. Therefore, $D$ implicitly identifies different versions of the surveillance database. We define the "observed epidemic curve" $C_D(t)$ as the number of symptomatic cases $C$ with symptom onset at date $t$ reported by the surveillance system at a date of reporting $D$. We define the "consolidated epidemic curve" $C^*(t)$ as the epidemic curve that is observed at the end of the outbreak, when all cases have been inserted in the database. For dates of symptom onset sufficiently distant in the past from $D$, we can assume that the observed epidemic curve is consolidated, i.e., that there is a time interval $\theta$ such that for $t \leq D - \theta$, $C_D(t)$ equals $C^*(t)$. We thus define for each reporting date $D$ a "consolidation function", $\pi_D$, representing the proportions of cases recorded in the consolidated curve $C^*$ and with symptom onset dates between $D - \theta$ and $D$ that were already reported at date $D$:

$$\pi_D(z) = \frac{C_D(D-z)}{C^*(D-z)}, (0 \leq z < \theta).$$

The consolidation function $\pi_D$ can be interpreted also as the proportion of cases having symptom onset at date $t = D - z$ that are reported in the surveillance database within $z$ days after symptom onset (i.e., at date $D$). By definition, $\pi_D$ cannot be known exactly at the date of reporting $D$; however, if we can obtain an approximation $\hat{\pi}_D$ using data available until $D$, an approximation $\hat{C}_D(t)$ of $C^*(t)$ can be inferred at date $D$:

$$C^*(t = D - z) \cong \hat{C}_D(t = D - z) = \frac{1}{\hat{\pi}_D(z)} C_D(t = D - z).$$

We refer to $\hat{C}_D(t)$ as the "nowcasted epidemic curve" estimated at time $D$.

From the approximated $\hat{\pi}_D$, we can additionally define a "consolidation delay" $T_{D,F}$, representing the minimum number of days that need to elapse before the number of cases of a given symptom onset date exceeds a given fraction $F$ of all cases that will be recorded at the end of the outbreak:

$$T_{D,F} = \min\{z : \hat{\pi}_D(z) \geq F\}$$

The consolidation delay gives an indication on until which symptom onset date the number of cases observed at time $D$ can be considered sufficiently complete; for example, we can consider the epidemic curve observed at time $D$ to be at least 90% complete until symptom onset date $t \leq D - T_{D,90}$.

Estimation of consolidation functions and delays

One way to estimate $\hat{\pi}_D(z)$ is to rely on consolidation functions relative to symptom onset dates $t \leq D - \theta$ that are consolidated at the date of reporting $D$. These consolidation functions can be obtained by comparing the number of cases for a given symptom onset date across successive reporting updates. For example, for $t = D - \theta$ we will have:

$$p_{D,0}(z) = \frac{C_{D-\theta+z}(D-\theta)}{C^*(D-\theta)} = \frac{C_{D-\theta+z}(D-\theta)}{C_D(D-\theta)}$$

since we have assumed that $C_D(t = D - \theta) = C^*(t = D - \theta)$. Equivalently, for any symptom onset date $t = D - \theta - i$, $i \geq 0$:

$$p_{D,i}(z) = \frac{C_{D-\theta-i+z}(D-\theta-i)}{C_D(D-\theta-i)}$$

We propose to approximate the unknown consolidation function $\pi_D(z)$ with the average over consolidation functions $p_{D,i}$ relative to the $N$ symptom onset dates closest to $D$ that are consolidated at $D$:

$$\hat{\pi}_D(z) = \frac{1}{N} \sum_{i=0}^{N-1} p_{D,i}(z)$$

Benchmarking the nowcasting method

We applied the proposed nowcasting method to data on confirmed symptomatic SARS-CoV-2 infections collected by regional health authorities in Italy and collated by the Istituto

Superiore di Sanità (Italian National Institute of Health) within the Italian COVID-19 integrated surveillance system [14] (a description of the system is reported in the Supplementary Material). Here, we consider the national-level epidemic curves by date of symptom onset as reported daily between May 1, 2020, and December 31, 2021. For practical purposes, we considered the epidemic curve observed on April 5, 2023 (i.e., 16 months after the end of the study period) as consolidated for symptom onset dates until December 31, 2021.

In the baseline analysis, we assumed $\theta$=30 days and chose $N$=30, so that the first reporting date for which the proposed method could be applied was June 29, 2020; in the Supplementary Material we provide a justification for these choices and sensitivity analyses of results obtained with alternative choices. For each reporting date, we estimated the corresponding approximate consolidation functions $\hat{\pi}_D(z)$, the nowcasted epidemic curves $\hat{C}_D$, and consolidation delays $T_{D,F}$ at different values of completeness ($F$ between 10% and 90% at intervals of 10%). We then estimated, for each reporting date, the values of the time-varying reproduction number computed from the observed epidemic curve ($R_D(t)$, or "net reproduction number"), and from the nowcasted one ($\hat{R}_D(t)$, or "nowcasted reproduction number"), evaluated at dates $D - T_{D,F}$, and we compared them with corresponding estimates from the consolidated epidemic curve ($R^*(t)$, or "reference reproduction number") (see Supplementary Materials for methodological details). For each level of completeness $F$, we denote with $\boldsymbol{R}_F$ and $\boldsymbol{\hat{R}}_F$ respectively the vectors of estimates $R_D(D - T_{D,F})$ and $\hat{R}_D(D - T_{D,F})$, obtained for different values of $D$.

A practical example

Figure 1 reports a practical example of the proposed method for a specific reporting date $\Delta$ = April 1, 2021, of the COVID-19 dataset. The closest date of symptom onset for which the number of cases can be assumed to be consolidated (i.e., numbers for this date will not change in reported epidemic curves successive to April 1, 2021) is $\Delta - \theta$ = March 2, 2021, given our choice of $\theta$=30 days. The estimate of $p_{\Delta,0}$ will be given by the number of cases with symptom onset on March 2 reported in epidemic curves produced between March 2 and April 1 and normalized by the consolidated number reported on April 1 (Figure 1a). The estimate of $p_{\Delta,N-1}$, for $N$=30, will be given by the number of cases with symptom onset on $\Delta - \theta - (N - 1)$ = February 1, reported between February 1 and March 3, normalized by the number reported on March 3. After averaging the $N$ observed consolidation functions, we obtain the consolidation function $\hat{\pi}_\Delta(z)$ relative to the date of reporting $\Delta$ (Figure 1b), from which we can obtain the corresponding consolidation delays $T_{\Delta,F}$ for different levels of completeness $F$, the nowcasted epidemic curve $\hat{C}_\Delta(t)$ (Figure 1c), and the estimates of the reproduction numbers $R_\Delta(t)$, $\hat{R}_\Delta(t)$, $R^*(t)$ (Figure 1d). This example shows that the net reproduction number significantly underestimated the reference value when the completeness values was below 90%. Thus, in this example the latest reliable estimate obtainable on $\Delta$ = April 1, 2021, using the net reproduction number is relative to $T_{\Delta,90}$= 16 days before the date of reporting (March 16, 2021).

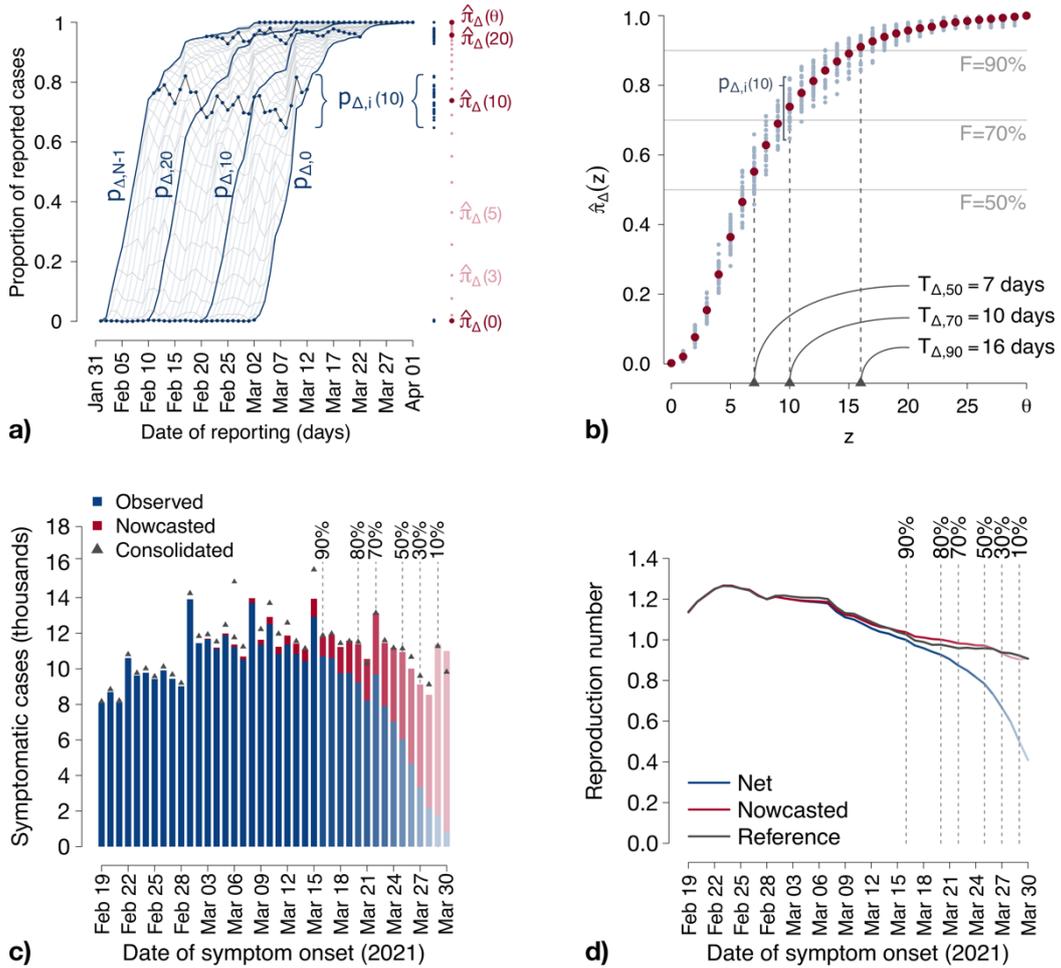

Figure 1. Example of application of the proposed nowcasting technique using data from the Italian COVID-19 integrated surveillance system for a selected reporting date ($\Delta$ = April 1, 2021). a) Approximation of the consolidation function. Light blue lines represent consolidation functions for symptom onset dates between February 1 ($p_{\Delta,N-1}$) and March 2 ($p_{\Delta,0}$), 2021; $p_{\Delta,0}, p_{\Delta,10}, p_{\Delta,20}$ and $p_{\Delta,N-1}$ are reported in thicker and darker lines to highlight the variability across consolidation functions. Dotted blue lines connect the values of different consolidation functions for the same number $z$ of days elapsed since the corresponding symptom onset, $p_{\Delta,i}(z)$, for $z$=0, 10, 20 and $\theta$ (gray lines show $p_{\Delta,i}(z)$ for the other values of $z$). The $p_{\Delta,i}(z)$ distributions for the selected values of $z$ are reported on the right-hand side of the panel as blue dots, and their mean, corresponding to the estimated value of $\hat{\pi}_\Delta(z)$, is reported further to the right as a dark red dot. Pink dots represent the mean of $p_{\Delta,i}(z)$ distributions, corresponding to the estimated value of $\hat{\pi}_\Delta(z)$, for the remaining values of $z$. b) Estimated consolidation function and corresponding consolidation delays. Light blue dots represent the distributions of $p_{\Delta,i}(z)$ for all values of $z$; dark red dots represent the estimated consolidation function at $\Delta$, $\hat{\pi}_\Delta(z)$. Horizontal lines define selected completeness thresholds (50%, 70% and 90%) and corresponding vertical lines define the corresponding consolidation delays. c) Observed and nowcasted epidemic curves by date of symptom onset. The consolidated epidemic curve is shown as dark gray triangles. Vertical dashed lines show the dates at which the observed number of cases is estimated to have reached a given completeness value. Bars in the epidemic curve are reported in fading colors with a level of darkness proportional to the estimated completeness. d) Mean estimates of the net, nowcasted, and reference reproduction numbers over time. Vertical dashed lines show the dates at which the observed number of cases is estimated to have reached a given completeness value. The level of darkness in line colors is proportional to the estimated completeness. Values for the reference reproduction number are shown as a dark gray line.

## Performance metrics

For each value of completeness $F$ and each of the $n$=551 considered reporting dates (between June 29, 2020 and December 31, 2021), we calculated: the absolute error of the net and nowcasted estimates of the reproduction number against the corresponding reference value; the proportion of all reporting dates in which each estimate underestimated the reference value; and the proportion of all reporting dates for which either estimate was closer than the other to the reference value.

Additionally, we defined an "epidemic period" as a sustained period of time during which the reproduction number remains above the epidemic threshold of 1. Specifically, we defined the start of an epidemic period as the first of at least 3 consecutive reporting days where the estimate of the reproduction number was above 1; and the end of an epidemic period as the day before the first of at least 7 consecutive reporting days where the estimate of the reproduction number was below 1. We compared the delays with which the net and nowcasted reproduction numbers were able to identify epidemic periods of duration longer than 15 days, as defined by the reference reproduction number.

## Results

During the study period, the Italian COVID-19 integrated surveillance system took about 6 days (median over the study period; 95% quantile: 5-8 days) from symptom onset to record at least 50% of all cases, and about 13 days (95% quantile: 7-17 days) to record at least 90% (Figure 2).

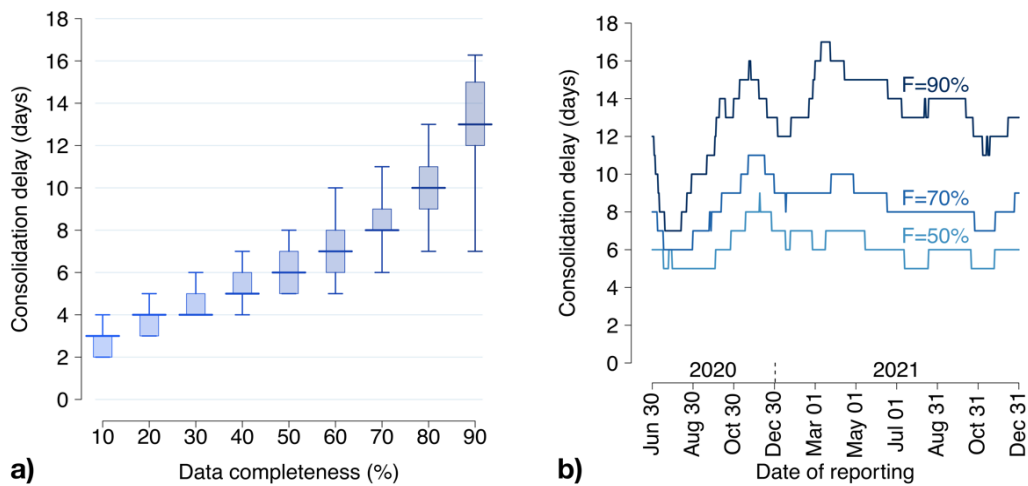

**Figure 2. Consolidation delays for the Italian COVID-19 integrated surveillance system.** a) Distribution of the consolidation delay across the estimation period (June 29, 2020 – December 31, 2021) for different values of completeness. Boxplots show the median (horizontal line), interquartile range (rectangle) and 95% quantiles (whiskers) over the n=551 reporting dates. b) Values at different reporting dates for selected values of completeness.

Overall, the net reproduction numbers evaluated at the date of 90% completeness, $R_{90}$, was a good approximation of the reference reproduction number $R^*$, resulting in a median absolute error over the entire study period of 0.042 (interquartile range, IQR: 0.025-0.066, Figure 3). The corresponding nowcasted estimate $\hat{R}_{90}$ had a median absolute error of 0.017 (IQR: 0.007-0.036, significantly smaller than the error for $R_{90}$: p-value of paired t-test between the net and nowcasted errors $\ll$ 0.001).

When choosing lower thresholds for completeness, the accuracy of the net reproduction number degraded rapidly, but not the one for nowcasted estimates. For example, with a completeness of 70% (corresponding to a median delay of 8 days), the median error was 0.116 (IQR: 0.078-0.171) for $R_{70}$ but 0.032 (IQR: 0.016-0.068) for $\hat{R}_{70}$, i.e., significantly smaller than the error for $R_{90}$ (paired t-test p-value << 0.001). The median error for $\hat{R}_{10}$ (0.115; IQR: 0.057-0.191) was comparable to the median error for $R_{70}$ (paired t-test p-value: 0.42). The net estimate systematically underestimated the reference value, while the nowcasted did so in about half of the reporting dates, as would be expected from an unbiased estimator; the latter result was independent on the considered completeness (Figure 3b). Furthermore, the nowcasted estimates were closer to the reference value, compared to net estimates obtained with the same completeness. This occurred for more than 90% of reporting dates when completeness values of 80% or lower were considered (Figure 3c).

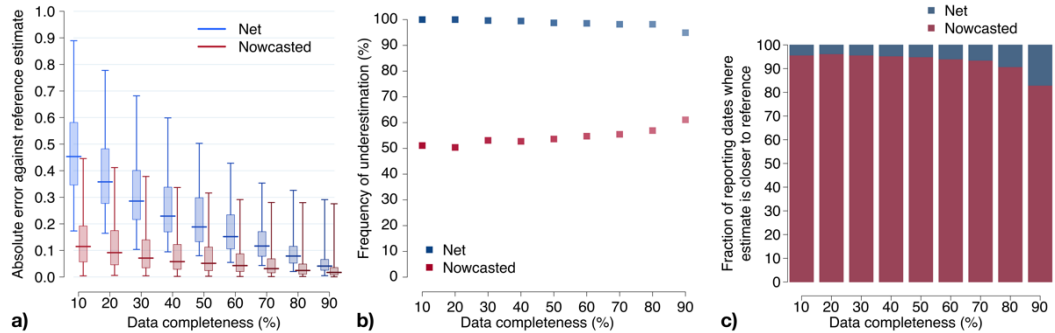

Figure 3. Accuracy of the net and nowcasted reproduction numbers. a) Distributions of the absolute error between the reference reproduction number and the net and nowcasted reproduction numbers, computed at different reporting dates (daily between June 29, 2020, and December 31, 2021), and evaluated at the date corresponding to a specified level of completeness. Boxplots show the median (horizontal line), interquartile range (rectangle) and 95% quantiles (whiskers). b) Fraction of reporting dates for which the estimate of the reproduction number underestimates the reference value. c) Fraction of reporting dates for which either estimate is the closest one to the reference value, for different values of completeness.

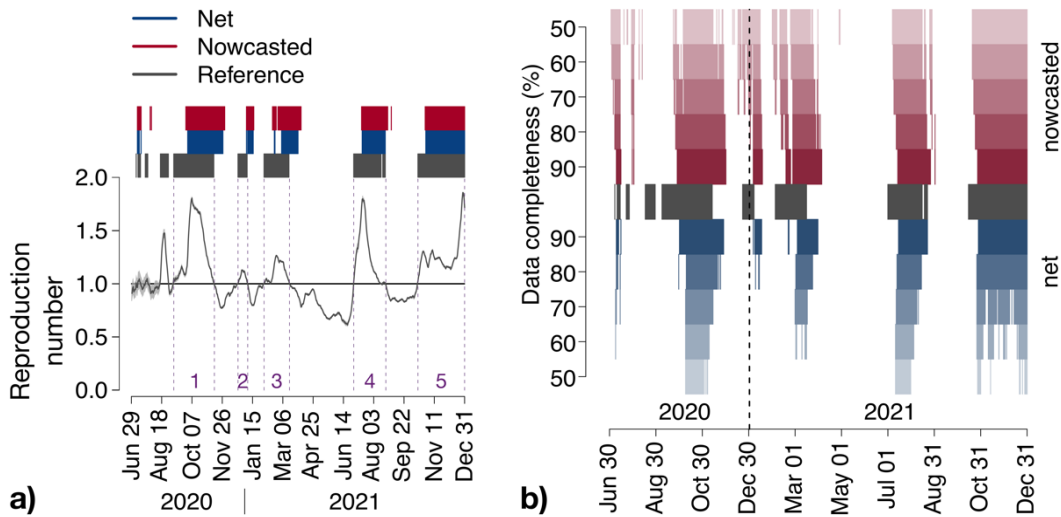

Figure 4. COVID-19 epidemic periods in Italy (2020-2021) and their detection. A) Reference reproduction number by date of symptom onset, as computed from consolidated data reported on April 5, 2023 (solid black line). The gray shaded area around the reference reproduction number (visible only in the period May-August 2020, due to the low number of cases contributing to R estimates) represents the 95%CI in its estimate. Dark gray bars above the graph highlight the days where the reference reproduction number was above 1, and purple vertical lines delimit epidemic periods (see Table 1), labeled with a progressive number just above the x axis. Blue and red bars identify the dates of reporting for which the net and nowcasted reproduction numbers, respectively,

computed at a completeness of 90% were above 1. B) Dates of reporting for which the net and nowcasted reproduction numbers computed at completeness between 50 and 90% were above 1.

We identified 5 epidemic periods where the reference reproduction number was above the epidemic threshold for more than 15 days between June 29, 2020, and December 31, 2021 (Figure 4 and Table 1). The net estimate at 90% completeness detected the epidemic periods with a delay that ranged between 13 and 30 days (Table 1). The net estimate at 70% completeness reduced by 2-3 days the detection delay for epidemic periods 4-5 but increased the delay by 8 days for period 1 and missed the detection of period 2. The net estimate at 50% completeness missed the detection of 3 periods out of 5. The nowcasted estimate allowed to anticipate the detection by 1 to 15 days (delays range: 12-20 days) at 90% completeness and by 6 to 23 days (delays range: 7-17 days) at 70% completeness compared to the net estimate at 90% completeness. In the case of 50% completeness, the nowcasted estimate performed worse than the net estimate at 90% completeness during period 1 (delay of 27 days vs. 23) but reduced the detection delay by 8 days for periods 4 and 5; for periods 2 and 3, the estimate provided an early warning 6 and 4 days compared to the actual start of the periods; however, it falsely flagged an additional epidemic period between July 2 and August 13, 2020.

| Epidemic period | | 1 | 2 | 3 | 4 | 5 |
|---|---|---|---|---|---|---|
| Date of start | | Sep 07, 2020 | Dec 22, 2020 | Feb 03, 2021 | Jul 01, 2021 | Oct 15, 2021 |
| Date of end | | Nov 12, 2020 | Jan 06, 2021 | Mar 16, 2021 | Aug 22, 2021 | After Dec 31, 2021 |
| Duration (days) | | 67 | 16 | 42 | 53 | >78 |
| Date of detection | Net (90% completeness) | Sep 30, 2020 | Jan 07, 2021 | Mar 05, 2021 | Jul 15, 2021 | Oct 28, 2021 |
| | Net (70% completeness) | Oct 8, 2020 | - | Mar 01, 2021 | Jul 12, 2021 | Oct 26, 2021 |
| | Net (50% completeness) | Oct 10, 2020 | - | - | Jul 11, 2021 | - |
| | Nowcasted (90% completeness) | Sep 27, 2020 | Jan 05, 2021 | Feb 17, 2021 | Jul 14, 2021 | Oct 27, 2021 |
| | Nowcasted (70% completeness) | Sep 24, 2020 | Dec 30, 2020 | Feb 10, 2021 | Jul 09, 2021 | Oct 22, 2021 |
| | Nowcasted (50% completeness) | Oct 4, 2020 | Dec 16, 2020 | Jan 30, 2021 | Jul 07, 2021 | Oct 20, 2021 |
| Detection delay (days) | Net (90% completeness) | 23 | 16 | 30 | 14 | 13 |
| | Net (70% completeness) | 31 | - | 26 | 11 | 11 |
| | Net (50% completeness) | 33 | - | - | 10 | - |
| | Nowcasted (90% completeness) | 20 | 14 | 14 | 13 | 12 |
| | Nowcasted (70% completeness) | 17 | 8 | 7 | 8 | 7 |
| | Nowcasted (50% completeness)* | 27 | -6 | -4 | 6 | 5 |

*The nowcasted estimate at 50% completeness falsely flagged an additional epidemic period (July 2 - August 13, 2020).

**Table 1. Characteristics of COVID-19 epidemic periods in Italy (2020-2021) and detection delays due to consolidation of epidemic curves.** The dates of start and end of epidemic periods refer to dates in which the reference reproduction number (computed from consolidated data reported on

April 5, 2023) was above 1 (see methods for the algorithmic definition). The dates of detection correspond to the first date of analysis for which the reproduction number, estimated at the date of the given completeness, was above 1.

## Discussion

We proposed a methodology that can improve situational awareness during an epidemic outbreak by nowcasting the reproduction number. Specifically, our methodology outperforms the traditional estimation of the time-varying reproduction number by improving both the accuracy and timeliness of the estimates. The only requirement of the method is having access to successive updates of epidemic curves by epidemiological surveillance systems, which are exploited to estimate consolidation delays and nowcast incomplete epidemic curves. Benchmarking the method against extensive data from over 18 months of the COVID-19 epidemics in Italy has demonstrated that its application can significantly improve the quality of the estimates. Compared to the same method based on the observed epidemic curve, the proposed nowcasting method: i) reduces the estimation delay for the COVID-19 reproduction number from a median of 13 to a median of 8 days while maintaining a similar accuracy; ii) halves the mean absolute error while maintaining the same estimation delay; or iii) allows improvements for both accuracy and timeliness of the estimates. More importantly, the nowcasted estimates markedly reduced (by between 6 and 23 days in the case of 70% completeness; with 50% completeness the nowcasted estimates did not provide clear improvements) the delay in detecting the beginning of epochs of sustained epidemic circulation, a notoriously difficult task for forecasting approaches [15].

In Italy, official estimates of the net reproduction number were obtained and made available to the public throughout the COVID-19 emergency with reference to 14 days before the date of reporting (approximately corresponding to a completeness of 90%). The analysis proposed here validates this choice, given the remarkable worsening of accuracy that would be obtained with lower accepted completeness although with shorter delays. Nowcasted estimates (based on a simplified and less extensively tested version of the proposed method) were used by national and regional health authorities throughout the course of the pandemic to gather further insights during weekly situation assessments, thus improving situational awareness.

There are a few limitations that need to be considered when interpreting our benchmarking results on SARS-CoV-2 data. First, the estimates of the reproduction numbers were obtained by assuming a fixed distribution of the generation time throughout the study period, corresponding to the one estimated for SARS-CoV-2 ancestral lineages. This choice was taken for simplicity and based on estimates for the Alpha [16], Delta [16] and Omicron variants [17] suggesting limited changes in the distribution of the generation time in Italy. Second, our main analysis compared different reproduction numbers only in terms of their mean estimate, without considering their variability. When the number of cases in epidemic curves is in the orders of the thousands, as for daily SARS-CoV-2 symptomatic cases in Italy, the estimated variability of the reproduction number is generally narrow, as shown in Figure 4a. Therefore, our results remained robust when considering alternative error functions that consider the variability in estimates (see Supplementary Materials). For what concerns the definition of epidemic periods, we used a heuristic definition to compare the potential in early warning of the net and nowcasted estimates at different levels of completeness. This definition does not distinguish situations of moderate (R slightly above 1) vs. catastrophic (R much above 1) epidemic growth and therefore does not necessarily correspond to the need for public health decision makers to take urgent action. Still, the identified epidemic periods corresponded to the main periods of expansion of the COVID-19 epidemics in Italy during the study period (Figure 4a), including the second wave in the fall of 2020 (period 1), the short resurgence during Christmas holidays of 2020 (period 2), the wave related to the expansion of the Alpha variant in spring 2021 (period 3), the increase of cases in summer 2021, partially related to the celebrations for the Italian victory in Euro2020 soccer championship [18] (period 4), and the wave due to Delta in the fall of 2021, replaced by Omicron in the last week of the year [19] (period 5). Although daily update of epidemic curves for SARS-CoV-2 were available until April 15, 2022 (after which updates were no longer done during the weekends), we decided to stop the benchmarking exercise on December 31, 2021, given the lower severity of the pandemic in 2022 [20,21,22], the broad diffusion of self-tests to be performed at home, and the progressive shift towards hospital surveillance. A future direction would be to test the methodology on COVID-19-associated hospitalizations as opposed to symptomatic cases.

The proposed method has potentially broad applications to infectious disease outbreaks and to different epidemiological endpoints (e.g., hospitalized cases). Although the concept of adjusting incidence curves is not new [8–13], it has never been extensively tested against the possibility of improving real-time estimates on the time-varying reproduction number and on the capacity of identifying periods of epidemic growth. A few conditions that may limit the use of our approach: first, it requires that a surveillance system is in place and provides regular updates of observed epidemic curves by date of symptom onset. This situation is not uncommon, especially for those outbreaks of diseases for which the real-time monitoring of reproduction number has highest importance. For example, regular situation reports containing epidemic curves by dates of symptom onset are usually issued by the World Health Organization during outbreaks of infectious diseases of concern such as Ebola disease, or by national and subnational authorities, for example during arboviral outbreaks in temperate climates. Second, the algorithm may be less accurate when applied to epidemic curves with limited numbers of cases: in these cases, consolidation functions may be highly noisy, impacting on the accuracy of nowcasting. In some cases, this problem may be potentially worked around by aggregating epidemic curves in appropriate units of time with respect to the length of the generation time. Third, the accuracy of the method may depend on the choice of its two parameters: the number of days after which we can consider a symptom onset date as consolidated ($\theta$) and the number of observed consolidation functions used to nowcast the epidemic curves ($N$). In the specific case of COVID-19 in Italy, the choice of $\theta$=30 days and $N$=30 was retrospectively shown to be appropriate in the Supplementary Material, where we additionally reported results of sensitivity analyses with respect to these values. However, when using the method in a new situation, how to appropriately choose $\theta$ and $N$ may not be obvious. The value of $\theta$ is itself dependent on the same data delays that we want to estimate, and represents an abstraction, as there is no theoretical time limit with which an epidemic curve may be retrospectively updated. An educated guess of an appropriate value can be given by observing the consolidation delays related to different symptom onset dates in the early phases of the epidemic under study. For what concerns the value of $N$, it should be sufficiently high to stabilize the variability of observable consolidation functions; however, the higher is $N$, the older is the information included in the estimation of consolidation functions. This is especially relevant in periods when consolidation delays are rapidly changing, due for example to improvements in the logistics of testing and reporting, or conversely to the overload of testing laboratories due to large increases in infection incidence. Another limit of the method is that the first nowcasted estimates are available at least $\theta + N - 1$ days after the first date of reporting available (another reason for preferring lower values in the choice of $N$).

All in all, the suggested limits of applicability imply that the proposed method might work best within outbreaks which are closely monitored by surveillance systems, with larger case numbers, and with a sufficiently long duration. These conditions, however, are exactly those where having a solid situational awareness and making informed decisions is of highest importance, such as during a pandemic.

Funding

This research was supported by EU funding within the NextGeneration EU-MUR PNRR Extended Partnership initiative on Emerging Infectious Diseases (Project no. PE00000007 INF-ACT) and by EU grant 874850 MOOD (catalogued as MOOD 000).

Conflict of Interest

MA received funding from Seqirus not related to the manuscript's subject. All other authors declare no conflicts of interest.

# Supplementary materials

# Increasing situational awareness through nowcasting of the epidemic curve and reproduction number

Andrea Bizzotto [a,b,*], Giorgio Guzzetta [a,*], Valentina Marziano [a], Martina del Manso [c], Alberto Mateo Urdiales [c], Daniele Petrone [c], Andrea Cannone [c], Chiara Sacco [c], Piero Poletti [a], Mattia Manica [a], Agnese Zardini [a], Filippo Trentini [d,e], Massimo Fabiani [c], Antonino Bella [c], Flavia Riccardo [c], Patrizio Pezzotti [c,#], Marco Ajelli [f,#], Stefano Merler [a,#]

[a] Center for Health Emergencies, Bruno Kessler Foundation, Trento, Italy
[b] University of Trento, Italy
[c] Department of Infectious Diseases, Istituto Superiore di Sanità, Rome, Italy
[d] Covid Crisis Lab, Bocconi University, Milan, Italy
[e] Department of Social and Political Sciences, Bocconi University, Milan, Italy
[f] Laboratory for Computational Epidemiology and Public Health, Department of Epidemiology and Biostatistics, Indiana University School of Public Health, Bloomington, Indiana, United States

[*] Joint first authors
[#] Joint senior authors

## 1. Integrated SARS-CoV-2 surveillance system

Since February 27, 2020, the Istituto Superiore di Sanità has coordinated a national case-based surveillance system reporting on all human cases with laboratory-confirmed SARS-CoV-2 infections [S1] defined as per the concurrent European Case Definition [S2]. Until January 2021, all cases were confirmed by RT-PCR; after that date, confirmation with rapid antigen tests was also accepted [S3]. The system was mandated by national law, regulated by dedicated technical documents that described data and data quality requirements and provided weekly data quality verification reports and ad hoc verification of incomplete/inconsistent items to each of the 21 Italian Regions and Autonomous Provinces [S4].

The system was based on an online dedicated and secure platform initially compiled manually by regional public health officers. As the epidemic progressed, with high case-loads, standardization for the upload of regional datasets was defined and implemented. The system allowed the collection of demographic data, geographic location data, date of symptom onset, date of diagnoses, date of hospitalization as well data on clinical severity and outcome. The definition of a case as symptomatic and of its date of symptom onset included any SARS-CoV-2-related symptoms, independently of its severity, as defined by clinical evaluation of the case. The definition of a case as imported was based on exposure outside the Italian territory ascertained through epidemiological investigations.

Results of the surveillance system were summarized in a daily dashboard, epidemiological bulletins and used as one of the sources of a mixed method risk assessment protocol that supported pandemic response in Italy [S5].

## 2. Estimation of the time-varying reproduction number

The estimation of the time-varying reproduction number was performed by applying a standard method [S6–8] which requires as input the number of autochthonous (locally transmitted) cases by date of symptom onset $A(t)$, the number of imported cases (infections acquired outside the geographical setting of interest) $I(t)$, and an estimate of the distribution of the generation time $\varphi(t)$ for the infection under study. The posterior distribution of $\tilde{R}(t)$ was obtained by applying Markov Chain Monte Carlo with Metropolis-Hasting sampling ($n = 50,000$ iterations) to the likelihood function defined below:

$$\mathcal{L} = \prod_{t=1}^{D} P\left(A(t); \tilde{R}(t) \sum_{s=1}^{t} \varphi(s) C(t-s)\right)$$

where:
- $P(k; \lambda)$ is the probability mass function of a Poisson distribution (i.e., the probability of observing k events if these events occur with rate λ).
- $C(t) = A(t) + I(t)$ is the total epidemic curve (total number of cases with symptom onset at time $t$).
- $\varphi(s)$ is the integral of the probability density function of the generation time evaluated between day s-1 and s.
- $\tilde{R}(t)$ is the daily reproduction number at time $t$.

The posterior distribution of the time-varying reproduction number $R(t)$ was computed by applying a weekly moving average to the posterior distribution of $\tilde{R}(t)$.

The same procedure reported above was applied to the consolidated epidemic curves $C^*(t)$ in order to estimate the reference reproduction number $R^*(t)$, and, at each reporting date $D$, to the net and nowcasted epidemic curves ($C_D(t)$ and $\hat{C}_D(t)$ respectively), in order to estimate the net ($R_D(t)$) and nowcasted ($\hat{R}_D(t)$) reproduction numbers over time.

For application to the Italian COVID-19 data, nowcasting was applied only to the autochthonous component of the epidemic curve, based on the observation that imported cases represented in most cases a negligible fraction of the total (median 0.4%, IQR 0.20-1.88% across symptom onset dates between January 28, 2020, and December 31, 2021). For the distribution of the generation time $\varphi(s)$, we used the distribution of the serial interval estimated in Italy for ancestral lineages [S9], given by a gamma function with shape 1.87 and rate 0.28, for a mean of 6.68 days. Further estimates of the distribution of the generation time in Italy for successively emerged variants showed minimal changes with respect to these values [S10, S11].

## 3. Retrospective assessment on the choice of θ

A critical assumption in the method for nowcasting epidemic curves is that, for $t \leq D - \theta$, the reported epidemic curve at day $D$, $C_D(t)$, approximates the consolidated epidemic curve $C^*(t)$. In order to assess the goodness of the chosen value for $\theta$ (equal to 30 days in the baseline analysis) when nowcasting COVID-19 epidemic curves in Italy, we computed the mean absolute percentage error $\eta(\theta)$ and the root mean squared error $\varepsilon(\theta)$ between the last values considered stable for the reported epidemic curve, given by $\tilde{C}(t;\theta) = C_{t+\theta}(t)$ at a certain value of $\theta$, and the corresponding reference values $C^*(t)$:

$$\eta(\theta) = \frac{1}{n}\sum_{t=t_0}^{T} \frac{|\tilde{C}(t;\theta) - C^*(t)|}{C^*(t)}$$

$$\varepsilon(\theta) = \sqrt{\frac{1}{n}\sum_{t=t_0}^{T} |\tilde{C}(t;\theta) - C^*(t)|^2}$$

Where $t_0$ and $T$ are the first and last symptom onset dates over which the error is computed and $n = T - t_0 + 1$ is the number of data points in $\tilde{C}(t)$. We evaluated $\eta(\theta)$ and $\varepsilon(\theta)$ for $\theta$ between 1 and 50, over the period comprised between $t_0$ = May 1, 2020 (the first reporting date that was available in our dataset) and $T$ = November 11, 2021 (the last date for which $\tilde{C}$ could be computed when $\theta$=50), resulting in $n$=560 data points. FigureS1 shows that both error metrics stabilize when θ is above 20 days, confirming that θ=30 days was an appropriate choice for Italian COVID-19 data.

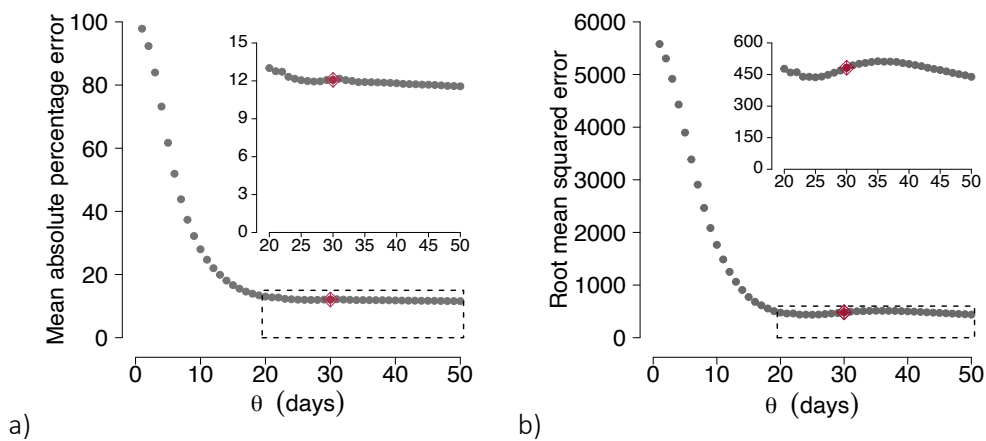

**Figure S1. Error metrics for assessing the choice of θ.** a) Mean absolute percentage error, $\eta(\theta)$; b) Root mean square error, $\varepsilon(\theta)$. The error value for the adopted value of θ=30 days is highlighted in red. To better appreciate differences in errors for values of θ>20 days, insets show the same graph with a zoom on the y-axis.

## 4. Sensitivity analysis on θ and N

The performance of the method proposed in this study may depend on the choice of its two parameters: the number of days after which a symptom onset date can be considered consolidated ($\theta$) and the number of observed consolidation functions used to nowcast the epidemic curves ($N$). In order to evaluate the sensitivity of model results with respect to the chosen values, we repeated the whole analysis for 6 additional combinations of $N$ and $\theta$, where both parameters can assume values of 20, 30 and 40. Since the first nowcasted estimates are available $\theta + N - 1$ days after the date of the first dataset, we evaluated results for reporting dates that were common to all combinations of $\theta$ and $N$, i.e. between July 19, 2020 (first date of availability of nowcasted estimates for $N = \theta = 40$) and December 31, 2021. Figures S2-S3 show minimal changes in the estimation of delays and in the performance of the method when choosing alternative values of the two parameters for selected data completeness levels ($F$ = 50%, 70% and 90%).

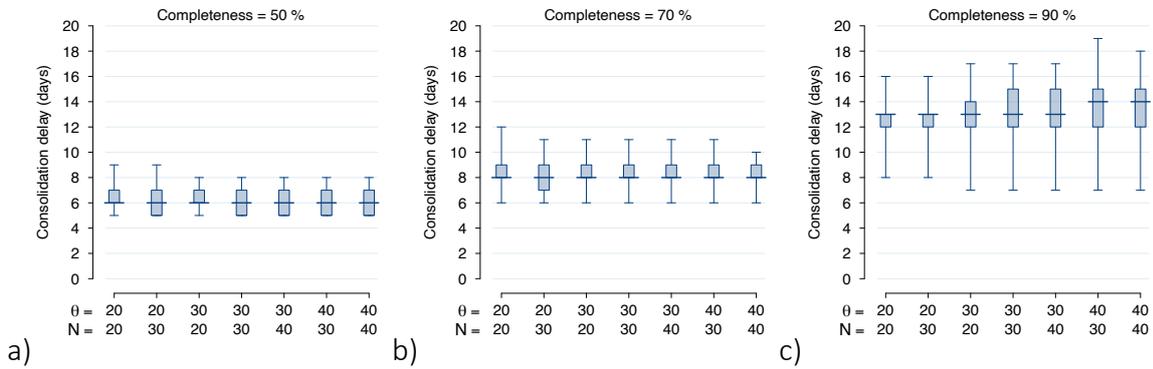

Figure S2. Distribution of consolidation delays for different combinations of the parameters $\theta$ and $N$. a) 50% completeness. B) 70% completeness. C) 90% completeness.

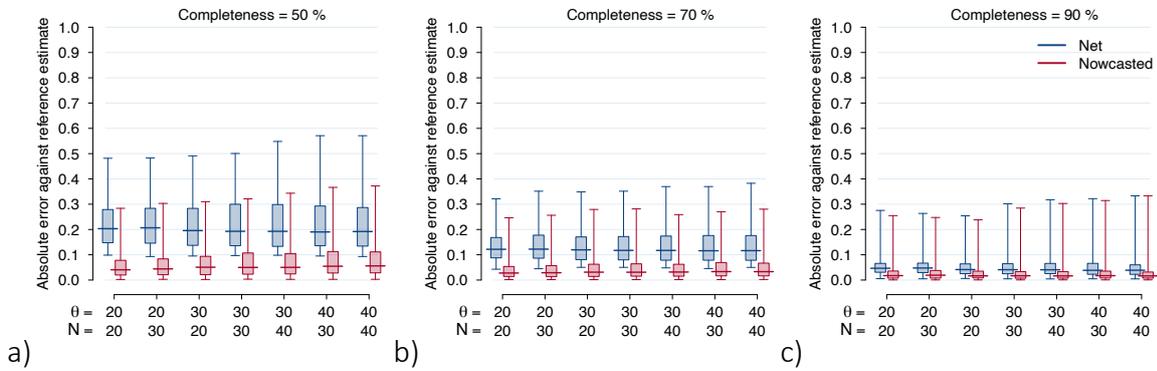

Figure S3. Distribution of absolute error functions for different combinations of the parameters $\theta$ and $N$. a) 50% completeness. B) 70% completeness. C) 90% completeness.

## 5. Further results with alternative error functions

In this section, we report results on further performance metrics for the net and nowcasted reproduction numbers in the baseline analysis ($N$=30, $\theta$=30 days).
Given the net reproduction numbers $R_F(D) = R(D - T_{D,F})$ estimated for a given data completeness $F$ and at each reporting date $D$ and consolidation delay $T_{D,F}$, we considered:
i) the bias error:

$$E_F^B(D) = R^*(D) - R_F(D),$$

ii) the percentage error:

$$E_F^P(D) = 100 \cdot \frac{|R_F(D) - R^*(D)|}{R^*(D)},$$

iii) the error with tolerance:
$$E_F^T(D) = \max\{R_F(D) - R^*_{97.5}(D); R^*_{2.5}(D) - R_F(D); 0\},$$

where $R^*_{2.5}$ and $R^*_{97.5}$ are respectively the lower and upper bounds of the 95%CI for the posterior distribution of $R^*$. The error with tolerance is defined in such a way to be zero when the average value of the net reproduction is within the 95% confidence interval of the reference estimate, and to be equal to the distance between the estimate and the closest boundary of the confidence interval otherwise.

The corresponding errors were computed for the nowcasted reproduction numbers by substituting $R_F$ with $\hat{R}_F$.

Figure S4 and Table S1 compare the overall performance of the net and nowcasted estimates for different values of data completeness. The bias error shows a significant downward bias (underestimation) for the net reproduction number and a substantial lack of bias for the nowcasted reproduction number at all levels of data completeness (see also Figure 3B in the main text). The percentage error shows that the net estimate at 90% completeness and the nowcasted estimate at 70% completeness approximate the reference value with an accuracy higher than 95% in a majority of reporting dates. The results of the error with tolerance may be better appreciated in Figure S5 and Table S2, where we considered the fraction of times in which the net and nowcasted estimates are:

i) correct (when the mean estimate falls within the 95%CI of the reference);
ii) underestimated (when the mean estimate falls below the 95%CI of the reference);
iii) overestimated (when the mean estimate falls above the 95%CI of the reference).

Net estimates of the reproduction number are correct only for 9% of reporting dates at a data completeness level of 90%, against 26% for the corresponding nowcasted estimate. Even with a data completeness of 50%, the nowcasted estimates are correct in a higher proportion of reporting dates (14%) compared to the net estimate at 90%.

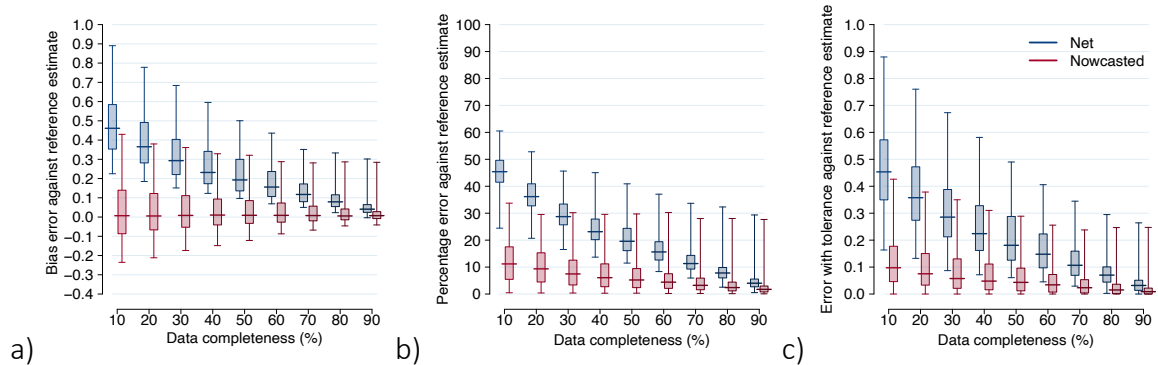

**Figure S4. Comparison of performance between the net and nowcasted estimates using alternative error functions.** a) Bias error; positive values represent underestimations of the reproduction number. B) Percentage error; c) Error with tolerance on the statistical variability of the reference estimate.

| Estimate | F (%) | Absolute error (median and IQR) | Bias error (median and IQR) | Percentage error (median and IQR) | Error with tolerance (median and IQR) |
|---|---|---|---|---|---|
| Nowcasted | 50 | 0.051 [0.023, 0.109] | 0.008 [-0.035, 0.084] | 5.3% [2.7, 9.8] | 0.044 [0.012, 0.097] |
| Nowcasted | 70 | 0.032 [0.016, 0.068] | 0.006 [-0.021, 0.055] | 3.4% [1.6, 6.2] | 0.023 [0.003, 0.054] |
| Nowcasted | 90 | 0.017 [0.007, 0.036] | 0.007 [-0.009, 0.028] | 1.8% [0.8, 3.1] | 0.009 [0, 0.022] |
| Net | 70 | 0.116 [0.078, 0.171] | 0.116 [0.078, 0.171] | 11.2% [9.1, 14.3] | 0.105 [0.067, 0.158] |
| Net | 90 | 0.042 [0.025, 0.066] | 0.039 [0.023, 0.064] | 4.0% [2.7, 5.6] | 0.029 [0.012, 0.049] |

Table S1. Comparison of performance metrics for the net and nowcasted estimates for selected levels of data completeness.

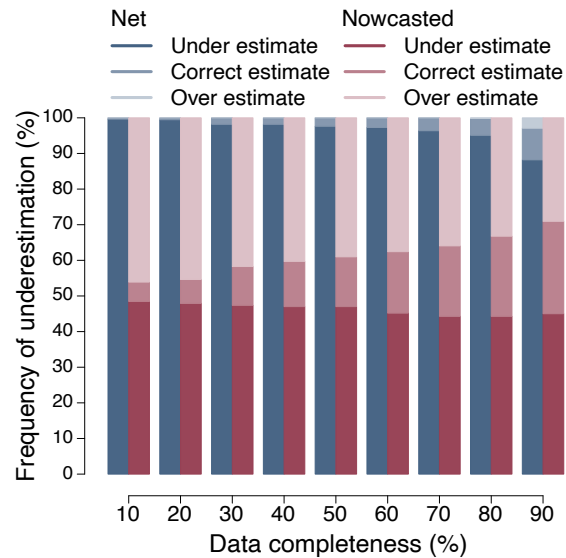

Figure S5. Performance of the net and nowcasted reproduction numbers when considering the stochastic variability of the reference estimate, for different data completeness levels.

| Method | Data completeness (%) | Under-estimate % | Over-estimate % | Correct estimate % |
|---|---|---|---|---|
| Nowcasted | 50 | 47 | 39 | 14 |
| Nowcasted | 70 | 44 | 36 | 20 |
| Nowcasted | 90 | 45 | 29 | 26 |
| Net | 90 | 88 | 3 | 9 |

Table S2. Performance of the net and nowcasted reproduction numbers when considering the stochastic variability of the reference estimate, for selected data completeness levels.

## 6. Alternative definition of the data consolidation delay

In the main text, we considered explicitly the case in which consolidation functions $\hat{\pi}_\Delta(z)$ are strictly monotonic with $z$ (Figure 1). However, consolidation functions in actual data may be non-monotonic when the process of consolidation due to reporting delays is counterbalanced by the retrospective removal of cases associated to a date of symptom onset. This may happen for a number of reasons, such as the re-evaluation of the date of symptom onset for a case after additional epidemiological investigations or after identification of data entry errors, the reclassification of cases as non-cases, or as asymptomatic cases, the identification of case duplications in the dataset, and so on. In these situations, the functions $\hat{\pi}_\Delta(z)$ may exceed the value of 1 at some $z$ between 0 and $\theta$. Figure S6 shows a comparison of the consolidation function for two reporting dates: July 15, 2020, and July 15, 2021. We evaluated the effect of an alternative definition of the data consolidation delay $T_{\Delta,F}$ where we considered, instead of the earliest $z$ for which $\hat{\pi}_\Delta(z)$ reaches the desired completeness $F$, the latest $z$ for which $\hat{\pi}_\Delta(z)$ is less than $1 - F$ away from its ideal value of 1:

$$T'_{\Delta,F} = \max_z \{|1 - \hat{\pi}_\Delta(z)| \leq 1 - F \ \wedge \ |1 - \hat{\pi}_\Delta(z-1)| > 1 - F \}.$$

An example of this second definition is shown in Figure S6 for July 15, 2020, and $F$=90%.
In Figure S7, differences in the distribution of the data consolidation delay when considering the alternative definition are shown for different values of $F$ and limited to the 210 reporting dates with a non-monotonic consolidation function. For $F$<70%, the data consolidation delays are identical; for $F$=80% and 90% the alternative definition produces longer data consolidation delays, but with substantially similar medians. In particular, with a data completeness of 80%, the alternative definition yields the same median delay of 10 days but with a 95% confidence interval shifted upward (7-22 days versus 6-13 days in the baseline); with a data completeness of 90% the median delay grows from 12 days (95%CI 7-15) in the baseline to 13 days (95%CI 8-26) in the alternative definition. Table S3 shows that the alternative definition improves the performance of both the net and nowcasted estimates at 80% and 90% completeness, although at the cost of higher delays.

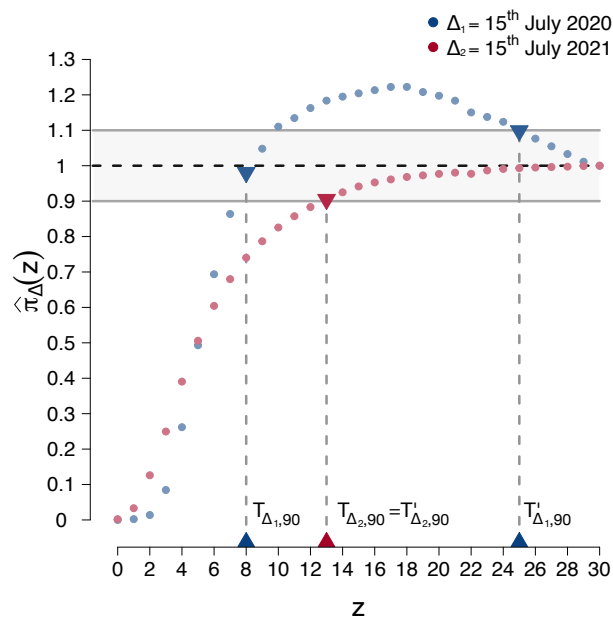

Figure S6. Example of alternative definitions for the data consolidation delay. Blue and red dots represent the consolidation functions associated to reporting dates July 15, 2020, and July 15, 2021, respectively. Triangles on the x-axis represent the estimated data consolidation delays according to the two definitions. In the case of a monotonic curve ($\Delta_2$ = July 15, 2021), the two definitions are equivalent, while when the curves are non-monotonic ($\Delta_1$ = July 15, 2020), the two definitions may give different results.

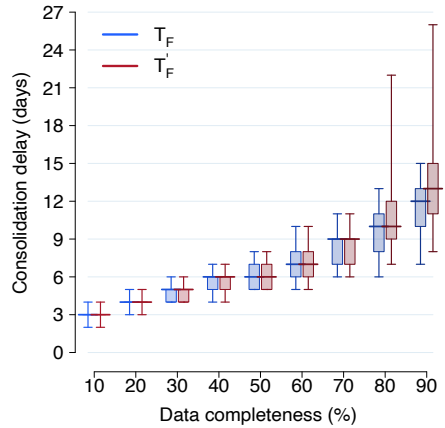

**Figure S7. Impact of alternative definitions for the data consolidation delay.** The figure shows the data consolidation delay distributions according to the baseline and alternative definition.

| Method | Definition of data consolidation delay | Absolute error (median and IQR) |
|---|---|---|
| Nowcasted | Baseline | 0.048 [0.02, 0.106] |
| Nowcasted | Alternative | 0.044 [0.019, 0.102] |
| Nowcasted | Baseline | 0.041 [0.018, 0.091] |
| Nowcasted | Alternative | 0.031 [0.016, 0.07] |
| Net | Baseline | 0.083 [0.055, 0.168] |
| Net | Alternative | 0.074 [0.041, 0.148] |
| Net | Baseline | 0.054 [0.027, 0.099] |
| Net | Alternative | 0.041 [0.018, 0.093] |

**Table S4.** Absolute errors of the nowcasted and net reproduction numbers against the reference estimate, using alternative definitions of the data consolidation delay, for data completeness levels of 80% and 90%.